\documentclass[twocolumn,
superscriptaddress,amsmath, amssymb, amsfonts,preprintnumbers,aps,prd,
longbibliography,nofootinbib]{revtex4-2}

\usepackage{scalerel}
\usepackage{color}
\usepackage{amsmath,amsfonts,amssymb}
\usepackage{slashed}
\usepackage{latexsym,epsfig}
\usepackage{dsfont}
\usepackage{arydshln}
\usepackage{extarrows}
\usepackage{hyperref}
\usepackage{array}

\newcommand{\sfrac}[2]{{\textstyle\frac{#1}{#2}}}

\def\moth{\mathsurround=0pt}
\newdimen\zo \zo=0pt

\def\tick{\leaders\hrule height 0.5ex depth 0pt \hskip 0.5pt}
\def\upboxfill{$\moth \setbox\zo\hbox{\tick}%
  \hskip 3pt\hbox to 0pt{$\tick$\hss}\hrulefill \hbox to 7.5pt{$\tick$\hss}$}

\def\dtick{\leaders\hrule height .34pt depth 0.5ex \hskip 0.5pt}
\def\downboxfill{$\moth \setbox\zo\hbox{\dtick}%
  \hskip 2pt\hbox to 0pt{$\dtick$\hss}\hrulefill \hbox to 2pt{$\dtick$\hss}$}

\def\ov{\bar}

\def\bec{\begin{center}}
\def\ec{\end{center}}

\def\k{\kappa}
\def\l{\lambda}

\def\m{\mu}
\def\n{\nu}
\def\r{\rho}
\def\s{\sigma}

\def\ov{\overline}

\def\nn{\nonumber}

\def\be{\begin{equation}}
\def\ee{\end{equation}}
\def\bea{\begin{eqnarray}}
\def\eea{\end{eqnarray}}
\def\ba{\begin{array}}
\def\ea{\end{array}}

\begin{document}

\title{From noncommutative Yang-Mills to noncommutative gravity \\ through a classical double copy map}

\author{Larisa Jonke} 
\email{larisa.jonke@irb.hr}
\affiliation{Division of Theoretical Physics, Rudjer Bo\v skovi\'c Institute, Bijeni\v cka 54, 10000 Zagreb, Croatia}
\affiliation{School of Theoretical Physics,  Dublin Institute for Advanced Studies,
 10 Burlington Road,  Dublin 4,  Ireland}

\author{Eric Lescano}
\email{eric.lescano@uwr.edu.pl}
\affiliation{University of Wroclaw, Faculty of Physics and Astronomy, Maksa Borna 9, 50-204 Wroclaw,
Poland}

\begin{abstract} 
We compute the first nontrivial noncommutative correction to the Einstein-Hilbert Lagrangian, which arises from the  double copy of noncommutative Yang-Mills theory (ncYM). We start by considering linear and quadratic $\theta$-corrections up to cubic order in fields in ncYM theory and in arbitrary $D$ dimensions. We compute the first nontrivial corrections to the three-points vertex operators
and use them to construct a double copy theory of the form ncYM $\times$ ncYM.  The resulting theory is given by a double geometrical formalism which includes noncommutative  corrections to the perturbative cubic double field theory (DFT) formulation, where the star product of the theory is doubled  in agreement with the doubling of the physical coordinates of the theory. Upon solving the level matching condition the noncommutative products are identified  and they produced $\theta^2$-corrections to the cubic DFT action. We analyze the pure gravitational limit of this formulation considering $D=4$ and imposing the transverse-traceless gauge. 
\noindent  

\end{abstract}

\maketitle

\section{Introduction}
Exploring the connections between gauge and gravitational theories has become a central focus of modern theoretical physics, towards the understanding of the quantum nature of gravitational interactions. One of the most successful programs in this line of research is the double copy prescription \cite{Dcopy1}-\cite{Dcopy3}, which provides a powerful framework linking scattering amplitudes of gravitational and gauge theories. First introduced by Kawai, Lewellen and Tye \cite{KLT} this approach reveals a remarkable correspondence in the context of string theory: closed string tree-level amplitudes can be recast in terms of open string amplitudes. Beyond its profound implications for scattering theory, the double copy formalism has also led to interesting connections between classical solutions in Yang-Mills theory and gravity \cite{class1}-\cite{class6}, playing an important role in the application of scattering amplitude methods to classical gravitational physics \cite{2class1}-\cite{2class8}.

Recently, the double copy formulation of Moyal-Weyl type noncommutative  gauge theories was discussed \cite{RDC1,RDC2}. There the twisted version of colour-kinematics duality compatible with the colour-kinematics mixing of noncommutative $U(N)$ gauge theory was developed. The result of the double copy relations in that case is the ordinary, commutative gravitational theory.  Using the classical double copy prescription, however, it is possible to capture the gravitational noncommutative corrections coming from the double copy map of noncommutative $SU(N)$ gauge theory. The explicit computation of these corrections, at their leading order in the noncommutativity parameter $\theta$, is the main goal of this letter.  

We exploit the idea of the classical double copy prescription which connects Yang-Mills theory with perturbative Double Field Theory (DFT) \cite{DFT}, the latter being a T-duality invariant reformulation of supergravity\footnote{For earlier works see \cite{Siegel1,Siegel2}.}. This prescription was given in \cite{HohmDC} for the  Yang-Mills theory expanded up to cubic order in fields,
\bea
\textrm{YM}^{(2,3)}(A) \rightarrow \textrm{DFT}^{(2,3)}(e) \rightarrow \textrm{Sugra}^{(2,3)}(h,b,\phi)~.
\label{Commutative}
\eea
While the  double copy map \eqref{Commutative} does not require a gauge fixing condition at quadratic order, the cubic contributions requires the Siegel gauge fixing at the DFT level, which is equivalent to the De Donder gauge once the level matching condition is solved. The same procedure has been used in other higher-derivative gauge theories for constructing T-duality invariant formulations \cite{LMR}-\cite{EA2}.  

We  start by considering linear and quadratic noncommutative corrections in $SU(N)$ gauge theory  as given in \cite{Jurco1}-\cite{Marija}, up to cubic order in fields.
Following \cite{HohmDC} but keeping track of the noncommutative structure of the theory in arbitrary $D$ dimensions, we compute the three-point vertex operators  $d_{abc}\rightarrow \pi^{(1)}_{\mu \nu \rho}$ and $f_{abc}\rightarrow \pi^{(0)}_{\mu \nu \rho} + \pi^{(2)}_{\mu \nu \rho}$.  
Using these operators, we construct the double copy ncYM $\times$ ncYM theory giving rise to a $\theta^2$-corrected theory defined in a double geometrical framework. This  is interpreted as a noncommutative correction to the cubic DFT Lagrangian (ncDFT) extending \eqref{Commutative}.

We study the pure gravitational limit ($b_{\mu \nu}=0$ and $\phi=0$) in the transverse-traceless gauge (gauge TT) solving the level matching condition in $D=4$. In this limit, the double copy theory is given by the Einstein-Hilbert Lagrangian plus Riemann cubic contributions, schematically,
\bea
\textrm{ncYM}^{(2,3)}  \rightarrow  &\textrm{DFT}^{(2,3)} +  \textrm{ncDFT}^{(3)}& \nn \\  & \downarrow \quad \quad \quad  \downarrow & \nn \\ & \textrm{GR}^{(2,3)} + \theta^2 \textrm{Riem}^3 & \, .
\eea
Thus we obtain the noncommutative corrections to pure gravity action from the double copy map.

This letter is organized as follows: In section \ref{review} we briefly review the classical double copy procedure given in \cite{HohmDC}. In section \ref{NC} we repeat the basic elements of the ncYM construction. In section \ref{action} we identify the vertex operators of ncYM to construct the double copy prescription. We then analyze the resulting  gravitational case in the transverse-traceless gauge. Finally in section \ref{Conclusions} we conclude our work including an outlook with promising future directions. 

\section{Double Field Theory as the double copy of Yang-Mills}
\label{review}
We start by reviewing the classical double copy map developed in~\cite{HohmDC}. We will briefly show how to obtain a quadratic and cubic double geometrical framework (which we understand as a T-duality invariant rewriting of the NSNS supergravity) using the double copy map. The  starting point is a $D$-dimensional Yang-Mills action,
\be
\label{YM_action}
S_{\rm{YM}} = - \frac{1}{4}\int d^{D}x\kappa_{ab}F_{\mu \nu}{}^{a}F^{\mu\nu b}\, ,
\ee
where 
\be
F_{\mu\nu}{}^{a} = 2\partial_{[\mu}A_{\nu]}{}^{a} + f^{a}{}_{bc}A_{\mu}{}^{b}A_{\nu}{}^{c}\, ,
\ee
and space-time indices are contracted with a Minkowski metric $\eta_{\mu\nu}={\rm{diag}}(-,+,+,+)$ and gauge indices are contracted by the Cartan-Killing metric $\k_{ab}$. The quadratic and cubic action require different maps, apart from the identification of the gauge field, so we will review them separately.

\subsection{Quadratic action}
 Passing  to momentum space, the quadratic terms of the gauge action read (up to a total derivative)
\be
\label{YM_momentum}
S_2 = - \frac{1}{2}\int_{k}\,\kappa_{ab}\,k^2 \,\Pi^{\mu\nu}(k) A_{\mu}{}^{a}(-k) A_{\nu}{}^{b}(k)\; ,  
\ee
where $\int_{k} \equiv \int d^D k$ and $A_\m{}^a(k)=(2\pi)^{-D/2}A_\m{}^a(x)e^{ikx}$. The projector $\Pi^{\mu\nu}(k)$ is defined as
\be
\label{PiProjector}
\Pi^{\mu\nu}(k) = \eta^{\mu\nu}-\frac{k^{\mu}k^{\nu}}{k^2}\, , 
\ee 
and obeys the identities  
\be
\label{projIdentity}
\Pi^{\mu\nu}(k)k_{\nu} = 0\;, \quad  \quad \Pi^{\mu\nu}\Pi_{\nu\rho} =  \Pi^{\mu}{}_{\rho}\; . 
\ee

The next step is to follow the double copy prescription which consists on replacing the colour indices by a second set of space-time indices ($a\rightarrow\bar{\mu}$) corresponding to a second set of space-time momenta $\bar{k}^{\bar{\mu}}$. This implies
\be
\label{DcFields}
A_{\mu}{}^{a}(k) \; \longrightarrow \; e_{\mu\bar{\mu}}(k,\bar{k})\;. 
\ee
The Cartan-Killing metric 
$\kappa_{ab}$ is identified following the relation
\be
\label{DCCartan}
\kappa_{ab} \; \longrightarrow \; \sfrac{1}{2}\, \bar{\Pi}^{\bar{\mu}\bar{\nu}}(\bar{k})\,, 
 \ee
where the projector $\bar{\Pi}^{\bar{\mu}\bar{\nu}}$ is defined in the same way as the projector $\Pi^{\mu\nu}$ but for barred momenta and indices instead of the original ones. Using these rules the quadratic action \eqref{YM_momentum} becomes
\be
\label{quadraticDoubleCopy}
S_2 = - \frac{1}{4}\int_{k,\bar{k}} k^2\,\Pi^{\mu\nu}(k)  \bar{\Pi}^{\bar{\mu}\bar{\nu}}(\bar{k}) e_{\mu\bar{\mu}}(-k,-\bar{k})e_{\nu\bar{\nu}}(k,\bar{k})\,.  
\ee
while the  level matching constraint states that 
\bea
\bar{k}^2=k^{2} \, .
\label{LMC}
\eea
The presence of a non-local term forces the introduction of an auxiliary scalar field $\phi(k,\bar k)$,
\be
\phi(k,\bar{k}) = \frac{1}{k^{2}}k^{\mu}\bar{k}_{\bar{\nu}}e_{\mu}{}^{\bar{\nu}} \, .
\ee
Considering the inclusion of the scalar field (dilaton), it is straightforward  to Fourier transform to a local action in doubled position space:
\begin{align}
\label{localDFTquadPosition}
S_2 = \frac{1}{4} & \int d^{D}x \,d^{D}\bar{x}\Big(e^{\mu\bar{\nu}}\square e_{\mu\bar{\nu}}+\partial^{\mu}e_{\mu\bar{\nu}}\,\partial^{\rho}e_{\rho}{}^{\bar{\nu}} \nn \\ 
& + \bar{\partial}^{\bar{\nu}}e_{\mu\bar{\nu}}\,\bar{\partial}^{\bar{\sigma}}e^{\mu}{}_{\bar{\sigma}} - \phi\square\phi + 2\phi\partial^{\mu}\bar{\partial}^{\bar{\nu}}e_{\mu\bar{\nu}}\Big)\, ,
\end{align}
which reproduces the standard quadratic DFT action \cite{DFT}. Pure gravity limit is obtained using  $x=\bar x$ as the solution of the level matching condition and identifying $e_{\m\n}$ with the linearized gravitational field $h_{\m\n}=g_{\m\n}-\eta_{\m\n}$, with $h:=h_\m^\m=\phi$.

\subsection{Cubic action}
After Fourier transforming to momentum space the cubic contributions from \eqref{YM_action}, the three-point vertex function with appropriate antisymmetrization is  obtained
\be
\label{Pi_3index}
\pi^{(0)\mu\nu\rho}(k_1,k_2,k_3) = \eta^{\mu\nu} k_{12}^{\rho} 
+ \eta^{\nu\rho} k_{23}^{\mu}
+ \eta^{\rho\mu} k_{31}^{\nu},
\ee
 for $k_{ij} = k_i - k_j$. The action at this point can be written as ($A_{i}\equiv A(k_{i})$)
\bea
S_3 &=& - \sfrac{i}{6\left(2\pi\right)^{D/2}}\int_{k_{i}}\delta\left(k_{1}+k_{2}+k_{3}\right)
\nonumber\\
&\times& f_{abc}\pi^{\mu\nu\rho}
A_{1\mu}^{a}A_{2\nu}^{b}A_{3\rho}^{c}\,,
\label{actionleading}
\eea
which shows that an extension of the double copy prescription \eqref{DcFields}, \eqref{DCCartan} must be considered in order to include the structure constant. The proper substitution rule is
\be
\label{DC_structure}
f_{abc} \longrightarrow \frac{i}{8} \bar{\pi}^{(0)\bar{\mu}\bar{\nu}\bar{\rho}}\, ,
\ee
with $\bar{\pi}^{(0)\bar{\mu}\bar{\nu}\bar{\rho}}$ defined in the same way as \eqref{Pi_3index} but for barred momenta. One obtains
\bea
\label{cubic}
S_3 & = & 
\frac{1}{48(2\pi)^{D/2}}\int dK_{1}dK_{2}dK_{3}
\delta(K_{1}+K_{2}+K_{3}) 
\nn \\
&\times& \bar{\pi}^{(0)\bar{\mu}\bar{\nu}\bar{\rho}}
\pi^{\rho\mu\nu}
e_{1\mu\bar{\mu}}e_{2\nu\bar{\nu}}e_{3\rho\bar{\rho}}
\eea
where $K = (k, \bar{k})$, $dK = d^{2D}K$ and $e_{i\mu\bar{\mu}} = e_{\mu\bar{\mu}}(K_i)$. After some manipulations, Fourier transformation to position space and integration by parts, the authors of \cite{HohmDC} obtained the following cubic action for the double copy of Yang-Mills theory
\bea
\label{YM_dc_cubic}
S_3 &=& \frac{1}{8}\int d^{D}x \,d^{D}\bar{x} \ e_{\mu\bar{\mu}}\Big[\,2\partial^{\m}e_{\r\ov\r}\,\ov \partial^{\ov\m}e^{\r\ov\r}-2\partial^{\m}e_{\n\ov\r}\,\ov \partial^{\ov\r}e^{\n\ov\m}\, 
\nn\\ 
&-& 2\partial^{\r}e^{\m\ov\r}\,\ov \partial^{\ov\m}e_{\r\ov\r} +\partial^{\r}e_{\r\ov\r}\,\ov \partial^{\ov\r}e^{\m\ov\m}+\ov \partial_{\ov\r}e^{\m\ov\r}\,\partial_{\r}e^{\r\ov\m}\, \Big].
\eea
This action agrees with the cubic (gauged fixed) DFT action. Upon solving the level matching condition (\ref{LMC}) using the solution $x=\bar x$, which breaks the double geometry, it gives rise to the universal NSNS sector of the low energy limit of string theory. In this scenario, we decompose the field $e_{\mu \nu}$ into a symmetric and an antisymmetric part,
\be
e_{\m\n}=h_{\m\n}+b_{\m\n}~ .
\ee
When both the dilaton (related to scalar $\phi$ appearing at quadratic order) and the $b$-field are set to zero, the resulting action reproduces linearized gravity in the De Donder gauge with $h=0$. In the next section, we  explore the noncommutative corrections to  action (\ref{YM_dc_cubic}) coming from the double copy map of ncYM  and imposing the transverse-traceless gauge, which is compatible with the De Donder gauge.   

\section{Noncommutative corrections}
\label{NC}
We consider the noncommutative extension of $SU(N)$ Yang-Mills theory using a Moyal-Weyl star product and considering terms up to cubic order in fields. The starting action is \cite{Jurco1}
\bea
S_{\textrm{YM}}^* = -\frac{1}{4} \textrm{Tr} \int d^{D}x  F^{*}_{\mu \nu} \star F^{* \mu \nu}  
\label{ncYM}
\eea
where
the Moyal-Weyl product is defined in the usual way
\begin{equation}
(f\star g)(x) = e^{\frac{i}{2} \theta^{\rho\sigma} \frac{\partial }{\partial x^\rho} \otimes
\frac{\partial }{\partial y^\sigma}}f(x) \otimes g(y)
\Big|_{y\rightarrow x} \, .
\end{equation}
Noncommutative field strength  is defined as 
\be
F^{*}_{\mu\nu} = 2\partial_{[\mu}A^{\star}_{\nu]} - i [A^{\star}_{\mu}, A^{\star}_{\nu}]_{\star}\, ,
\ee
where $A^\star$ is a solution of the Seiberg-Witten map \cite{SW}. Namely, the algebra of noncommutative gauge transformation in general  closes in enveloping algebra, and one introduces the map relating noncommutative fields and parameters with the standard ones in order to keep the same number of degrees of freedom \cite{NC1}. 
The $SU(N)$ generators $T_{a}$ satisfy
\bea
[T_{a},T_{b}]= i f_{ab}{}^{c} T_{c} \, , \\
\{T_{a},T_{b}\}= \frac{1}{N} \delta_{ab} + d_{ab}{}^{c} T_{c} \, ,
\eea
while the trace of three generators is given by
\bea
\textrm{Tr}(T_{a} T_{b} T_{c}) = \frac14(d_{abc}+if_{abc}) \, .
\label{trace}
\eea

The quadratic form of the action (\ref{ncYM}) is given by
\bea
S_2^* = - \int d^{D}x \partial_{[\mu} A_{\nu] a} \star \partial^{\mu} A^{\nu a} \, .
\eea
The $\theta$-corrections coming from this action are total derivatives, and therefore there are no noncommutative corrections at quadratic level in fields.

\subsection{$\theta$-correction to the cubic Yang-Mills action}
At cubic order, the $\theta$-correction to the Yang-Mills action is given by \cite{Jurco1}
\bea
S_3^* =  - \frac{1}{8}  \textrm{Tr} \int d^{D}x \, \theta^{\mu \nu}&\Big(& \{\{F_{\rho \mu},F_{\sigma \nu}\},F^{\rho \sigma}\} \nn \\
&& - \frac12 \{F_{\mu \nu},F_{\rho \sigma} F^{\rho \sigma}\} \Big).  
\eea
Using the expression (\ref{trace}) we rewrite this as
\bea
S_3^* = - \frac{1}{8}\int d^{D}x\, \theta^{\mu \nu} d_{a b c}  &\Big(&F_{\rho \mu}{}^{a} F_{\sigma \nu}{}^{b} F^{\rho \sigma}{}^{c} - \frac14 F_{\mu \nu}{}^{a} F_{\rho \sigma}{}^{b} F^{\rho \sigma}{}^{c} \Big). \nn
\eea
Moving to momentum space and regrouping terms we can rewrite the noncommutative contribution to the cubic action in terms of the three vertex operator $\pi^{(1)}$,
\bea
\frac{i}{6(2\pi)^{\frac{D}{2}}} \int_{k} \delta(k_{1}+k_{2}+k_{3}) d_{a b c} \pi^{(1) \mu \nu \rho} A_{1\mu}^{a} A_{2\nu}^{b} A_{3\rho}^{c} \, ,  \label{28}
\eea
where 
\bea
&&\pi^{(1)\mu \nu \rho} = \frac{3}{4}\Big(-\theta^{(\mu| \sigma} k_{1 \sigma} (k_{2}^{\nu} k_{3}^{\rho)} + k_{2 \epsilon} k_{3}^{\epsilon} \eta^{|\nu \rho)}) \nn \\
&&  + 2 \theta^{(\mu|\sigma} k_{1 \epsilon} k_{2 \sigma} k_{3}^{\epsilon} \eta^{|\nu \rho)} - 2 \theta^{\epsilon \sigma} k_{1\epsilon} k_{2\sigma} k_{3}^{(\mu} \eta^{\nu \rho)} \Big ) \, . 
\eea
Following the YM$\times$YM structure of the double copy prescription we identify 
\bea
d_{abc} \rightarrow \frac{ip}{8} \bar \pi^{(1) \bar \mu \bar \nu \bar \rho} \, ,
\eea
with $p$ a free parameter of order 1.
The above identification give rise to $\theta^2$- corrections to the cubic DFT action, which takes the implicit form 
\bea
-\frac{p}{48(2\pi)^{\frac{D}{2}}} \int_{k,\bar k} \delta(K_1+K_2+K_3) \bar \pi^{(1)\bar \mu \bar \nu \bar \rho} \pi^{(1)}_{\mu \nu \rho} e_1^{\mu}{}_{\bar \mu} e_2^{\nu}{}_{\bar \nu} e_3^{\rho}{}_{\bar \rho} \, , \nn\\
\eea
where now $\bar \pi^{(1)\bar \mu \bar \nu \bar \rho}$ is proportional to a second noncommutative parameter $\bar \theta$ coming from a $\bar \star$ product defined as
\begin{equation}
(f \bar \star g)(x) = e^{\frac{i}{2} \bar \theta^{\bar \rho \bar \sigma} \frac{\bar \partial }{\bar \partial \bar x^{\bar \rho}} \otimes
\frac{\bar \partial }{\bar \partial \bar y^{\bar \sigma}}}f(\bar x) \otimes g(\bar y)
\Big|_{\bar y\rightarrow \bar x} \, .
\end{equation}
We present the full noncommutative cubic action in  section \ref{action}, after taking into account the quadratic $\theta$-correction to the cubic Yang-Mills action.

\subsection{$\theta^2$-correction to the cubic Yang-Mills action}
The $\theta^2$-correction to the cubic part of the Yang-Mills action is given by \cite{Marija,Lutz}
\bea\label{nc3}
S^{*}_3= - \frac{1}{32} \int d^Dx &&f_{a b c} \theta^{\mu \nu} \theta^{\kappa \lambda}    \Big( \frac14 F_{\mu \nu}{}^{a} D_{\kappa} F_{\rho \sigma}{}^{b} D_{\lambda} F^{\rho \sigma c} \nn \\ &&\hspace{1.1cm} -D_{\mu} F_{\rho \kappa}{}^{a} D_{\nu}F_{\sigma \lambda}{}^{b} F^{\rho \sigma c}\Big)  \, , 
\eea
where $D_{\mu}=\partial_\m-i A_\mu$ is the standard gauge covariant derivative. The first term in the action \eqref{nc3} can be eliminated by using the covariant field redefinition $A_\s \rightarrow  A_\s+c\theta^{\mu \nu} \theta^{\kappa \lambda}D_\l F_{\m\n} F_{\k\s}$ with appropriately\footnote{Note that the results for the $\theta^2$-correction to the  cubic part of the Yang-Mills action obtained from Refs.\cite{Lutz,Marija} differ by a factor in front of the term  we removed by field redefinition. In \eqref{nc3} we used  result of Ref.\cite{Marija}.} chosen constant $c$, thus we drop it from further considerations.  

Moving to momentum space and regrouping terms we rewrite the action \eqref{nc3} in terms of the three-vertex operator $\pi^{(2)}$,
\bea
S^{*}_3=  \frac{i}{6(2\pi)^{\frac{D}{2}}} \int_{k} \delta(k_{1}+k_{2}+k_{3}) f_{a b c} \pi^{(2) \mu \nu \rho} A_{1\mu}{}^{a} A_{2\nu}{}^{b} A_{3\rho}{}^{c} \, ,   \label{squareactionYM} \nn \\
\eea
where the $\theta^2$-correction to (\ref{Pi_3index}) is given by
\bea
\pi^{(2)\mu \nu \rho} = \frac{3}{8} \theta^{\epsilon \xi}k_{1 \epsilon}k_{2 \xi}  \Big(   \theta^{[\mu \nu} k_{1 \kappa}  k_{2}^{\rho]} k_{3}^{\kappa} - \theta^{\kappa [\nu} k_{1 \kappa}  k_{2}^{\rho} k_{3}^{\mu]} \Big) \, . \nn \\
\eea
Thus we obtain two extra sources of the $\theta^2$-corrections to the DFT action: on one hand, we use the identification $f_{abc}\rightarrow \frac{i}{8} \pi^{(0)\bar \mu \bar \nu \bar \rho}$ on (\ref{squareactionYM}). On the other hand, we consider the identification
\bea
f_{abc}\rightarrow \frac{i}{8} \pi^{(2)\bar \mu \bar \nu \bar \rho}
\eea
in the leading order action (\ref{actionleading}). 
Therefore, the full identification for the structure constant $f_{abc}$ is given by
\bea
f_{abc}\rightarrow \frac{i}{8} (\pi^{(0)\bar \mu \bar \nu \bar \rho} + \pi^{(2)\bar \mu \bar \nu \bar \rho}) \, .
\eea

\section{Full $\theta^2$-corrected DFT action and its pure gravitational limit}
\label{action}
The full $\theta^2$-correction, obtained through the double copy map, to the cubic DFT action is given by
\bea\label{fullk}
&&S^{*}_3|_{\textrm{DFT}} = -\frac{1}{48(2\pi)^{\frac{D}{2}}} \int_{k,\bar k} \delta(K_1+K_2+K_3) \times\nn\\
&&\Big(p\bar \pi^{(1)}_{\bar \mu \bar \nu \bar \rho} \pi^{(1) \mu \nu \rho} 
+ \bar \pi^{(0)}_{\bar \mu \bar \nu \bar \rho} \pi^{(2) \mu \nu \rho}  
+ \bar \pi^{(2)}_{\bar \mu \bar \nu \bar \rho} \pi^{(0) \mu \nu \rho}\Big) e_{1\mu}{}^{\bar \mu} e_{2\nu}{}^{\bar \nu} e_{3\rho}{}^{\bar \rho} \, .\nn\\ 
\eea
The  action \eqref{fullk} contain terms proportional to $\bar \theta \theta$, $\theta^2$,  and $\bar \theta^2$. 
The explicit expression in coordinate space is given in (\ref{NCDFTaction}) of the appendix \ref{DFTaction}. 

\subsection{Pure gravitational analysis}
While the double copy map presented in this letter is valid in an arbitrary number of dimensions, from now on we will fix $D=4$. We solve the level matching constraint as $x=\bar x$ and we consider a pure gravitational setup $e_{\m\n}=h_{\m\n}$ satisfying the gauge TT, i.e., we impose $\partial^{\mu}h_{\mu \nu}=0=h$. Then double copy action up to cubic order in fields can be written schematically\footnote{The full expression is long and not very illuminating thus we do not present it at this level.} as
\bea
S^{*}_{(2,3)} =  -2 \int_{x}  h_{\m\n} R^{\mu \nu} + \int_{x} \, {\cal L}_{Riem^3} + \textrm{t.d} \ ,
\label{fullthetaaction}
\eea
where ${\cal L}_{Riem^3}$ are terms cubic in the Riemann tensor and t.d are total derivative terms containing $ \textrm{Riem}^2$ contributions, which drop out in agreement with the absence of nontrivial corrections at the quadratic level. 

All the cubic contributions in ${\cal L}_{Riem^3}$ proportional to the Ricci tensor can be eliminated considering a field redefinition of perturbation of the metric, $h_{\mu \nu} \rightarrow h^{*}_{\mu \nu}= h_{\mu \nu} + \tilde h_{\mu \nu}$ in the quadratic action i.e.,  
\bea\label{GR2GF}
& &  \int d^Dx~h^{*}_{\m\n} R^{\mu \nu}(h^*) \nn \\ \nn &=&  \int d^Dx~h_{\m\n} R^{\mu \nu}+ \int d^Dx~\tilde h_{\m\n} R^{\mu \nu} + \mathcal{O}(\theta^4)  \, .
\eea
The $\tilde h_{\m\n}$ is proportional to $\theta^2$Riem$^2$ and all the Ricci contributions of (\ref{fullthetaaction}) can be eliminated. The procedure induces new quartic contributions (in fields and $\theta$) through the commutative cubic action which now should depend on $h^{*}_{\mu \nu}$. Finally, the first nontrivial noncommutative corrections to  gravity action inferred from the linearized expressions are 
\bea\label{Riem3}
{\cal L}_{Riem^3} && = \theta^{\mu \nu} \theta^{\rho \sigma} \Big(\frac{p}{144} R^{\xi \alpha \beta \kappa} R_{\xi \alpha \beta \kappa} R_{\mu \nu \rho \sigma}  \nn \\ && + \frac{5p}{18} R^{\xi \beta \alpha \kappa} R_{\xi \alpha \beta \mu} R_{\kappa \nu \rho \sigma} - \frac{2p}{9} R^{\xi \alpha \beta \kappa} R_{\xi \alpha \beta \mu} R_{\kappa \rho \nu \sigma} \nn \\ && + (\frac{2}{9} p + \frac{1}{24}) R^{\xi \beta \alpha \kappa} R_{\xi \mu \alpha \rho} R_{\beta \nu \kappa \sigma}  \nn \\ && - (\frac{5}{18} p - \frac{1}{24}) R^{\xi \beta \alpha \kappa} R_{\xi \mu \alpha \rho} R_{\beta \sigma \kappa \nu}  \nn \\ && - \frac{p}{18} R^{\delta \beta \xi}{}_{\mu} R_{\delta \rho \xi}{}^{\kappa} R_{\beta \sigma \kappa \nu} + \frac{1}{12} R^{\delta \beta \alpha}{}_{\mu} R_{\delta}{}^{\kappa}{}_{\alpha \rho} R_{\beta \sigma \kappa \nu}\Big)  \, . \nn \\
\eea
 Note that the proposed noncommutative corrections to Einstein-Hilbert action should be thought of as higher derivative/curvature corrections obtained through the double copy map. These corrections could potentially have a geometric interpretation of the action in terms of $\star$-deformed general relativity \cite{NC3}. 

The equation of motion of the pure gravitational setup coming from the double copy of the noncommutative Yang-Mills Lagrangian is given by
\bea
&& \Box h^{\mu \nu} = - \frac18 \partial^{\nu} h^{\sigma \lambda} \partial^{\mu} h_{\lambda \sigma} - \frac14 \partial^{\sigma} h_{\lambda}{}^{\nu} \partial_{\sigma} h^{\lambda \mu}  + \frac14 \partial_{\sigma} h_{\lambda}{}^{\nu} \partial^{\lambda} h^{\mu \sigma} \nn \\ && - \frac18 \eta^{\mu \nu} \partial_{\lambda} h^{\sigma \rho} \partial_{\rho} h^{\lambda}{}_{\sigma} 
+ \frac{3}{16} \eta^{\mu \nu} \partial_{\lambda}h^{\rho \sigma} \partial^{\lambda} h_{\rho \sigma}
+ \frac12 C^{\mu \nu}(\theta^2) \,  ,
\label{eom}
\eea
where $C^{\mu \nu}(\theta^2)$ represents  $\theta^2$-corrections coming from the variation of the higher-derivative contributions in the double copy Lagrangian.  In order to check the possibility of propagation of unphysical degrees of freedom in this particular setup, we split the noncommutative corrections into cubic and quartic order in derivatives:
\bea
C^{\mu \nu}(\theta^2) = C_{1}^{\mu \nu}(\partial^3 h) + C_{2}^{\mu \nu}(\partial^4 h) \, .
\eea
The explicit form of the corrections is given the appendix \ref{eoms}. 
 One can show that if we restrict  noncommutativity to spatial directions only, there are no time derivatives higher than quadratic and therefore, no propagating unphysical degrees of freedom. Limiting nocommutativity to spacial direction\footnote{The Lorentz invariance is already broken with the choice of the constant noncommutativity parameter $\theta$. } is the standard solution for the potential loss of unitarity due to higher derivative corrections coming from the $\star$-product deformations. 

The present construction brings new possibilities when studying the noncommutative correction to exact gravitational solutions like the generalized version of the Kerr-Schild ansatz \cite{KS}, which has been constructed in \cite{KL1} and further extended in \cite{KL2}-\cite{EAKS2}. Since our double copy procedure generates a nontrivial cubic correction to the cubic DFT action, it is natural to expect $\theta^2$-corrections to the single and zeroth copy procedure following \cite{KL1} and using the prescription given in \cite{ET}.

\section{Conclusions and outlook}
\label{Conclusions}
We performed a classical double copy procedure on the noncommutative formulation of non-Abelian Yang-Mills theory up to cubic order in fields. We constructed the three-point vertex operators in order to identify the noncommutative corrections to the totally symmetric, $d_{abc}$, and totally antisymmetric, $f_{abc}$, structure constants and we use them to construct the double copy theory respecting the form ncYM$\times$ncYM.  The resulting theory (\ref{NCDFTaction}) can be interpret as a noncommutative deformation of the DFT cubic action\footnote{See \cite{ET} for a nonperturbative formulation of noncommutative DFT in generalized metric formalism.}. 

Interestingly enough, the noncommutativity is given by a pair of $\star$-products, with parameters $\theta^{\mu \nu}$ and $\bar \theta^{\bar \mu \bar \nu}$, in agreement with the doubling of the physical coordinates of DFT, also noticed in \cite{RDC1, RDC2}. As a consequence of solving the level matching condition (identification of the dual coordinates with the ordinary ones), these products are identified, giving rise to $\theta^2$-corrections to the cubic DFT action, while the quadratic part remains uncorrected. We wrote the covariant form of the noncommutative cubic correction to the action in the pure gravitational limit ($b_{\mu \nu}=0$, $\phi=0$) and using the transverse-traceless gauge in $D=4$, compatible with the De Donder gauge, required by the cubic commutative action. 

We computed the corrections to the equation of motion up to quadratic order in fields, giving rise to new corrections to the subleading dispersion relation for gravitational waves coming from the double copy formulation. In the present version of the construction the equation of motion potentially contains cubic and quartic time derivatives for generic $\theta$, which indicates that the foundation of the double copy requires the inclusion of extra degrees of freedom, possible in the form of charged scalar fields mimicking the technique developed in \cite{EA}. However, in the special case when $\theta$ is a spacial constant, we evade this issue. 
 
Furthermore, we use the transverse-traceless gauge in order to simply the first noncommutative correction to the Einstein-Hilbert Lagrangian, given by $\textrm{Riem}^3$ terms. Using this gauge, for example, all the cubic contributions containing the Ricci scalar (for instance, $ \textrm{Riem}R^2$) are trivial due to $R=0 + \mathcal{O}(h^2)$ in the gauge TT. Then, our result can be generalized by imposing only the De Donder gauge and allowing new gravitational contributions related to the Ricci scalar. Using the De Donder gauge one finds $R= -\partial^{\mu \nu}h_{\mu \nu} + \mathcal{O}(h^2)$, and only the leading order part contributes to the cubic action. In our case, the transverse condition eliminates the leading order contribution to $R$. 

 While the $\pi^{(0)}$ and $\pi^{(2)}$ 
 vertices are fixed using the perturbative DFT, the inclusion of the vertex $\pi^{(1)}$ induces a free parameter $p$ which cannot be fixed by the classical double copy procedure. The full cubic action, up to $\theta^2$-contribution can be written as
\bea
S^{*}_3 & = &  \frac{i}{6(2\pi)^{\frac{D}{2}}} \int_{k} \delta(k_{1}+k_{2}+k_{3})  A_{1\mu}{}^{a} A_{2\nu}{}^{b} A_{3\rho}{}^{c}  \nn \\ && [f_{a b c}(\pi^{(0) \mu \nu \rho} +\pi^{(2) \mu \nu \rho} )+d_{abc}\pi^{(1) \mu \nu \rho}]\, ,   \label{totalcubic} 
\eea
and due to different symmetrization of $f$ and $d$ terms we can write 
\bea
S^{*}_3 & = &  \frac{i}{6(2\pi)^{\frac{D}{2}}} \int_{k} \delta(k_{1}+k_{2}+k_{3})  A_{1\mu}{}^{a} A_{2\nu}{}^{b} A_{3\rho}{}^{c}  \nn \\ &&  (f_{a b c}+d_{abc})(\pi^{(0) \mu \nu \rho}  +\pi^{(1) \mu \nu \rho}+\pi^{(2) \mu \nu \rho})\, .   \label{totalcubic2} 
\eea
This rewriting implies the identification
\be
f_{a b c} + d_{abc}\to \sfrac{i}{8}(\pi^{(0) \mu \nu \rho}  +\pi^{(1) \mu \nu \rho}+ \pi^{(2) \mu \nu \rho}) \, ,
\ee
which fixes $p=1$. This choice is reminiscent of the identification encountered in the analysis of  double copy   of $U(N)$ noncommutative gauge theory based on twisted colour-kinematics duality \cite{RDC2}. It would be interesting to further study the colour-kinematics mixing inherent in the noncommutative field theories in the framework of (possibly weakly) constraint double field theory. 

 Another interesting continuation of our current results is to explore the b-field contributions to the Einstein-Hilbert action coming from the double copy map. These contributions will be proportional to $\theta^2$ after solving the level matching condition and we expect some contributions of the form $(\nabla H)^3$ after writing the terms in covariant form. The inclusion of the b-field to the noncommutative correction of the cubic action will allow us to study non-trivial T-duality transformations involving the metric and the b-field on the noncommutative cubic dynamics.

Finally, once the cubic corrections are fixed one can use purely algebraic methods to construct the quartic contributions, as described in Refs.\cite{olafnew, Zeitlin} in commutative setting and in \cite{RDC1} in noncommutative generalization of the problem. These become important 
in order to preserve causality when studying the group/phase velocity of the (modified) gravitational waves for propagating signals.  As studied in \cite{Maldacena}, the contributions to the group/phase velocity require an infinite expansion in derivatives in the gravitational Lagrangian, as happens in string theory. Interestingly, the double copy of ncYM offers a tower of higher-derivative corrections, unlike its commutative version.

\subsection*{Acknowledgments}
We kindly thank A. Chatzistavrakidis, M. Dimitrijevi\'c \'Ciri\'c and J. A. Rodriguez for discussion and feedback on the first version of the paper. E.L. thanks IAFE, in particular C. Nunez and D. Marques, for providing a workplace during the final stage of this project. E.L is supported by the SONATA BIS grant 2021/42/E/ST2/00304 from the National Science Centre (NCN), Poland. L.J. is supported by   the Croatian Science Foundation project “HigSSinGG — Higher Structures and Symmetries in Gauge and Gravity Theories” (IP-2024-05-7921) and by the European Union — NextGenerationEU. This article is also based upon work from the COST Action 21109 CaLISTA, supported by COST (European Cooperation in Science and Technology).

\appendix
\begin{widetext}
\section{Noncommutative double field theory: corrections to the cubic action}
\label{DFTaction}
The full noncommutative  $\theta^2$-correction to the perturbative DFT action obtained from the double copy map of ncYM is given by\footnote{We used CADABRA software \cite{Cadabra} for the calculations. \\ Code is available upon request.}
\bea
S^{*}_{3} & = & -\frac{3p}{256} \int_{x,\bar x} \Big( \theta^{(\mu|\sigma} \bar \theta_{(\bar \mu|\bar \sigma} \partial_{\sigma} \bar \partial^{\bar \sigma} e_{\mu}{}^{\bar \mu} \partial^{\nu} \bar \partial_{\bar \nu} e_{\nu}{}^{\bar \nu} \partial^{\rho)} \bar \partial_{\bar \rho)} e_{\rho}{}^{\bar \rho} 
+ \theta^{(\mu|\sigma} \bar \theta_{(\bar \mu|\bar \sigma} \partial_{\sigma} \bar \partial^{\bar \sigma} e_{\mu}{}^{\bar \mu} \partial^{\nu} \bar \partial^{\bar \epsilon} e_{\nu}{}^{\bar \nu} \eta_{|\bar \nu \bar \rho)} \partial^{\rho)} \bar \partial_{\bar \epsilon} e_{\rho}{}^{\bar \rho} \nn \\ &&
-2 \theta^{(\mu|\sigma} \bar \theta_{(\bar \mu|\bar \sigma} \partial_{\sigma} \bar \partial_{\bar \epsilon} e_{\mu}{}^{\bar \mu} \partial^{\nu} \bar \partial_{\bar \sigma} e_{\nu}{}^{\bar \nu} \partial^{\rho)} \bar \partial^{\bar \epsilon} e_{\rho}{}^{\bar \rho} \eta_{|\bar \nu \bar \rho)} 
+2 \theta^{(\mu|\sigma} \bar \theta^{\bar \epsilon \bar \sigma} \partial_{\sigma} \bar \partial_{\bar \epsilon} e_{\mu}{}^{\bar \mu} \partial^{\nu} \bar \partial_{\bar \sigma} e_{\nu}{}^{\bar \nu} \partial^{\rho)} \bar \partial_{(\bar \mu} e_{\rho}{}^{\bar \rho} \eta_{|\bar \nu \bar \rho)} \nn \\ &&
+ \theta^{(\mu|\sigma} \bar \theta_{(\bar \mu|\bar \sigma} \partial_{\sigma} \bar \partial^{\bar \sigma} e_{\mu}{}^{\bar \mu} \partial_{\epsilon} \bar \partial_{\bar \nu} e_{\nu}{}^{\bar \nu} \partial^{\epsilon} \bar \partial_{\bar \rho)} e_{\rho}{}^{\bar \rho} \eta^{|\nu \rho)} 
+ \theta^{(\mu|\sigma} \eta^{|\nu \rho)} \bar \theta^{(\bar \mu| \bar \sigma} \eta^{|\bar \nu \bar \rho)} \partial_{\sigma} \bar \partial_{\bar \sigma} e_{\mu \bar \mu} \partial_{\epsilon} \bar \partial_{\bar \epsilon} e_{\nu \bar \nu} \partial^{\epsilon} \bar \partial^{\bar \epsilon} e_{\rho \bar \rho} \nn \\ &&
-2 \theta^{(\mu|\sigma} \eta^{|\nu \rho)} \bar \theta_{(\bar \mu}{}^{\bar \sigma} \eta_{|\bar \nu \bar \rho)} \partial_{\sigma} \bar \partial_{\bar \epsilon} e_{\mu}{}^{\bar \mu} \partial_{\epsilon} \bar \partial_{\bar \sigma} e_{\nu}{}^{\bar \nu} \partial^{\epsilon} \bar \partial^{\bar \epsilon} e_{\rho}{}^{\bar \rho} 
+2 \theta^{(\mu|\sigma} \eta^{|\nu \rho)} \bar \theta^{\bar \epsilon \bar \sigma} \partial_{\sigma} \bar \partial_{\bar \epsilon} e_{\mu}{}^{\bar \mu} \partial_{\epsilon} \bar \partial_{\bar \sigma} e_{\nu}{}^{\bar \nu} \partial^{\epsilon} \bar \partial_{(\bar \mu} e_{\rho}{}^{\bar \rho} \eta_{|\bar \nu \bar \rho)} \nn \\ &&
-2 \theta^{(\mu|\sigma} \eta^{|\nu \rho)} \bar \theta_{(\bar \mu|\bar \sigma}  \partial_{\epsilon} \bar \partial^{\bar \sigma} e_{\mu}{}^{\bar \mu} \partial_{\sigma} \bar \partial_{\bar \nu} e_{\nu}{}^{\bar \nu} \partial^{\epsilon} \bar \partial_{\bar \rho)} e_{\rho}{}^{\bar \rho} 
-2 \theta^{(\mu|\sigma} \eta^{|\nu \rho)} \bar \theta_{(\bar \mu| \bar \sigma} \eta_{|\bar \nu \bar \rho)} \partial_{\epsilon} \bar \partial^{\bar \sigma} e_{\mu}{}^{\bar \mu} \partial_{\sigma} \bar \partial_{\bar \epsilon} e_{\nu}{}^{\bar \nu} \partial^{\epsilon} \bar \partial^{\bar \epsilon} e_{\rho}{}^{\bar \rho} \nn \\ &&
+4 \theta^{(\mu|\sigma} \eta^{|\nu \rho)} \bar \theta_{(\bar \mu| \bar \sigma} \eta_{|\bar \nu \bar \rho)} \partial_{\epsilon} \bar \partial_{\bar \epsilon} e_{\mu}{}^{\bar \mu} \partial_{\sigma} \bar \partial_{\bar \sigma} e_{\nu}{}^{\bar \nu} \partial^{\epsilon} \bar \partial^{\bar \epsilon} e_{\rho}{}^{\bar \rho}
-4 \theta^{(\mu|\sigma} \eta^{|\nu \rho)} \bar \theta^{\bar \epsilon \bar \sigma}  \partial_{\epsilon} \bar \partial_{\bar \epsilon} e_{\mu}{}^{\bar \mu} \partial_{\sigma} \bar \partial_{\bar \sigma} e_{\nu}{}^{\bar \nu} \partial^{\epsilon} \bar \partial_{(\bar \mu|} e_{\rho}{}^{\bar \rho} \eta_{|\bar \nu \bar \rho)} \nn \\ &&
+ 2 \theta^{\epsilon \sigma} \bar \theta_{(\bar \mu|\bar \sigma} \partial_{\epsilon} \bar \partial_{\bar \sigma} e_{\mu}{}^{\bar \mu} \partial_{\sigma} \bar \partial_{\bar \nu} e_{\nu}{}^{\bar \nu} \partial^{(\mu|} \bar \partial_{\bar \rho)} e_{\rho}{}^{\bar \rho} \eta^{|\nu \rho)}
+ 2 \theta^{\epsilon \sigma} \bar \theta_{(\bar \mu|\bar \sigma} \eta_{|\bar \nu \bar \rho)} \partial_{\epsilon} \bar \partial^{\bar \sigma} e_{\mu}{}^{\bar \mu} \partial_{\sigma} \bar \partial_{\bar \epsilon} e_{\nu}{}^{\bar \nu} \partial^{(\mu|} \bar \partial^{\bar \epsilon} e_{\rho}{}^{\bar \rho} \eta^{|\nu \rho)} \nn \\ &&
-4 \theta^{\epsilon \sigma} \bar \theta_{(\bar \mu|\bar \sigma} \eta_{|\bar \nu \bar \rho)} \partial_{\epsilon} \bar \partial_{\bar \epsilon} e_{\mu}{}^{\bar \mu} \partial_{\sigma} \bar \partial^{\bar \sigma} e_{\nu}{}^{\bar \nu} \partial^{(\mu|} \bar \partial^{\bar \epsilon} e_{\rho}{}^{\bar \rho} \eta^{|\nu \rho)}
+4 \theta^{\epsilon \sigma} \bar \theta_{\bar \epsilon \bar \sigma}  \partial_{\epsilon} \bar \partial_{\bar \epsilon} e_{\mu}{}^{\bar \mu} \partial_{\sigma} \bar \partial_{\bar \sigma} e_{\nu}{}^{\bar \nu} \partial^{(\mu|} \bar \partial_{(\bar \mu|} e_{\rho}{}^{\bar \rho} \eta^{|\nu \rho)} \eta_{|\bar \nu \bar \rho)} \Big)\nn \\  && 
+ \frac{1}{128} \int_{x, \bar x} \bar \theta^{\bar \epsilon \bar \xi} \bar \theta^{[\bar \mu \bar \nu}\Big(\partial_{\rho} \bar \partial_{\bar \epsilon} \bar \partial_{\bar \lambda} e_{\mu \bar \mu} \bar \partial_{\bar \xi} \bar \partial^{\bar \rho]} e^{\mu}{}_{\bar \nu} \bar \partial^{\bar \lambda} e^{\rho}{}_{\bar \rho} - \partial_{\mu} \bar \partial_{\bar \epsilon} \bar \partial_{\bar \lambda} e_{\rho \bar \mu} \bar \partial_{\bar \xi} \bar \partial^{\bar \rho]} e^{\mu}{}_{\bar \nu} \bar \partial^{\bar \lambda} e^{\rho}{}_{\bar \rho}  - \bar \partial_{\bar \epsilon} \bar \partial_{\bar \lambda} e_{\mu \bar \mu} \partial^{\rho} \bar \partial_{\bar \xi} \bar \partial^{\bar \rho]} e^{\mu}{}_{\bar \nu} \bar \partial^{\bar \lambda} e_{\rho \bar \rho} \nn \\ && + \bar \partial_{\bar \epsilon} \bar \partial_{\bar \lambda} e_{\mu \bar \mu} \partial^{\mu} \bar \partial_{\bar \xi} \bar \partial^{\bar \rho]} e^{\rho}{}_{\bar \nu} \bar \partial^{\bar \lambda} e_{\rho \bar \rho} - \bar \partial_{\bar \epsilon} \bar \partial_{\bar \lambda} e^{\mu}{}_{\bar \mu} \bar \partial_{\bar \xi} \bar \partial^{\bar \rho]} e^{\rho}{}_{\bar \nu} \partial_{\mu} \bar \partial^{\bar \lambda} e_{\rho \bar \rho} + \bar \partial_{\bar \epsilon} \bar \partial_{\bar \lambda} e^{\mu}{}_{\bar \mu} \bar \partial_{\bar \xi} \bar \partial^{\bar \rho]} e^{\rho}{}_{\bar \nu} \partial_{\mu} \bar \partial^{\bar \lambda} e_{\rho \bar \rho} \Big) \nn \\ && 
+ \frac{1}{128} \int_{x, \bar x} \bar \theta^{\bar \epsilon \bar \xi} \bar \theta^{\bar \kappa [\bar \nu} \Big( - \partial^{\rho} \bar \partial_{\bar \epsilon} \bar \partial_{\bar \kappa} e^{\mu}{}_{\bar \mu} \bar \partial_{\bar \xi} \bar \partial^{\bar \rho} e_{\mu \bar \nu} \bar \partial^{\bar \mu]}e_{\rho \bar \rho} + \partial^{\mu} \bar \partial_{\bar \epsilon} \bar \partial_{\bar \kappa} e^{\rho}{}_{\bar \mu} \bar \partial_{\bar \xi} \bar \partial^{\bar \rho} e_{\mu \bar \nu} \bar \partial^{\bar \mu]}e_{\rho \bar \rho} + \bar \partial_{\bar \epsilon} \bar \partial_{\bar \kappa} e_{\mu \bar \mu} \partial^{\rho} \bar \partial_{\bar \xi} \bar \partial^{\bar \rho} e^{\mu}{}_{\bar \nu} \bar \partial^{\bar \mu]}e_{\rho \bar \rho} \nn \\ && - \bar \partial_{\bar \epsilon} \bar \partial_{\bar \kappa} e_{\mu \bar \mu} \partial^{\mu} \bar \partial_{\bar \xi} \bar \partial^{\bar \rho} e^{\rho}{}_{\bar \nu} \bar \partial^{\bar \mu]}e_{\rho \bar \rho} + \bar \partial_{\bar \epsilon} \bar \partial_{\bar \kappa} e^{\mu}{}_{\bar \mu} \bar \partial_{\bar \xi} \bar \partial^{\bar \rho} e^{\rho}{}_{\bar \nu} \partial_{\mu} \bar \partial^{\bar \mu]}e_{\rho \bar \rho} - \bar \partial_{\bar \epsilon} \bar \partial_{\bar \kappa} e^{\rho}{}_{\bar \mu} \bar \partial_{\bar \xi} \bar \partial^{\bar \rho} e^{\mu}{}_{\bar \nu} \partial_{\mu} \bar \partial^{\bar \mu]}e_{\rho \bar \rho} \Big) \nn \\  && 
+ \frac{1}{128} \int_{x, \bar x} \theta^{ \epsilon  \xi} \theta^{[ \mu  \nu}\Big(\bar \partial^{\bar \rho}  \partial_{ \epsilon}  \partial_{ \lambda} e_{\mu  \mu}  \partial_{ \xi}  \partial^{ \rho]} e_{\nu}{}_{\bar \mu}  \partial^{ \lambda} e_{\rho \bar \rho} - \bar \partial^{\bar \nu}  \partial_{ \epsilon}  \partial_{ \lambda} e_{\mu  \bar \mu}  \partial_{ \xi}  \partial^{ \rho]} e_{\nu \bar \nu}  \partial^{ \lambda} e_{\rho}{}^{\bar \mu}  -  \partial_{ \epsilon}  \partial_{ \lambda} e_{\mu \bar \mu} \bar \partial^{\bar \rho}  \partial_{ \xi}  \partial^{ \rho]} e_{\nu}{}_{ \nu}  \partial^{ \lambda} e_{\rho  \bar \rho} \nn \\ && +  \partial_{ \epsilon}  \partial_{ \lambda} e_{\mu \bar \mu} \bar \partial^{\bar \mu}  \partial_{ \xi}  \partial^{ \rho]} e_{\nu}{}^{ \bar \rho}  \partial^{ \lambda} e_{\rho  \bar \rho} -  \partial_{ \epsilon}  \partial_{ \lambda} e_{\mu \bar \mu}  \partial_{ \xi}  \partial^{ \rho]} e_{\nu \bar \nu} \bar \partial^{\bar \mu}  \partial^{ \lambda} e_{\rho}{}^{\bar \nu} +  \partial_{ \epsilon}  \partial_{ \lambda} e_{\mu \bar \mu}  \partial_{ \xi}  \partial^{ \rho]} e_{\nu \bar \nu} \bar \partial^{\bar \nu}  \partial^{ \lambda} e_{\rho}{}^{\bar \mu} \Big) \nn \\ && 
+ \frac{1}{128} \int_{x, \bar x} \theta^{ \epsilon  \xi} \theta^{ \kappa [ \nu} \Big( - \bar \partial^{\bar \rho}  \partial_{ \epsilon}  \partial_{ \kappa} e_{\mu \bar \mu}  \partial_{ \xi}  \partial^{ \rho} e_{\nu}{}^{\bar \mu}  \partial^{\mu]}e_{\rho \bar \rho} + \bar \partial^{\bar \nu}  \partial_{ \epsilon}  \partial_{ \kappa} e_{\mu}{}^{\bar \rho}  \partial_{ \xi}  \partial^{ \rho} e_{\nu \bar \nu}  \partial^{ \mu]}e_{\rho \bar \rho} +  \partial_{ \epsilon}  \partial_{ \kappa} e_{\mu \bar \mu} \bar \partial^{\bar \rho}  \partial_{ \xi}  \partial^{ \rho} e_{\nu}{}_{\bar \mu}  \partial^{ \mu]}e_{\rho \bar \rho} \nn \\ && -  \partial_{ \epsilon}  \partial_{ \kappa} e_{\mu \bar \mu} \bar \partial^{\bar \mu}  \partial_{ \xi}  \partial^{ \rho} e_{\nu \bar \nu}  \partial^{ \mu]}e_{\rho}{}^{\bar \nu} +  \partial_{ \epsilon}  \partial_{ \kappa} e_{\mu \bar \mu}  \partial_{\xi} \partial^{ \rho} e_{\nu \bar \nu} \bar \partial^{\bar \mu}  \partial^{\mu]}e_{\rho}{}^{\bar \nu} -  \partial_{ \epsilon}  \partial_{ \kappa} e_{\mu \bar \mu}  \partial_{\xi} \partial^{\rho} e_{\nu \bar \nu} \bar \partial^{\bar \nu}  \partial^{ \mu]}e_{\rho  \rho} \Big) \, .
\label{NCDFTaction}
\eea

\section{Noncommutative gravity: corrections to the equations of motion}
\label{eoms}
Using the leading order equation of motion ($\Box h^{\mu \nu} + O(h^2)=0$) and field redefinitions, we simplify $C_{1}^{\mu \nu}$ and $C_{2}^{\mu \nu}$  to the form 
\bea
C_1^{\mu \nu} & = & \frac{7p}{48} \theta^{\mu \beta} \theta^{\nu \kappa} \partial^{\sigma \gamma}{}_{\beta}{h^{\delta \alpha}} \partial_{\delta \alpha \kappa}{h_{\sigma \gamma}} - ( \frac{1}{36} p + \frac{1}{768}) \theta^{\nu \alpha} \theta^{\beta \kappa} \partial^{\mu \tau}{}_{\alpha}{h^{\epsilon \xi}} \partial_{\epsilon \xi \beta}{h_{\tau \kappa}}  \nn \\ && + \frac{p}{36} \theta^{\nu \beta} \theta^{\alpha \kappa} \partial^{\mu \tau}{}_{\alpha}{h^{\epsilon \xi}} \partial_{\epsilon \beta \kappa}{h_{\tau \xi}} - (\frac{1}{768} p + \frac{1}{384}) \theta^{\nu \beta} \theta^{\alpha \kappa} \partial^{\mu \tau}{}_{\alpha}{h^{\epsilon \xi}} \partial_{\epsilon \xi \beta}{h_{\tau \kappa}}  \nn \\ && - \frac{p}{144} \theta^{\nu \beta} \theta^{\alpha \kappa} \partial^{\mu \tau \epsilon}h^{\xi}{}_{\alpha} \partial_{\xi \beta \kappa}{h_{\tau \epsilon}} - \frac{p}{48} \theta^{\nu \beta} \theta^{\alpha \kappa} \partial^{\tau \epsilon}{}_{\alpha}{h^{\mu \xi}} \partial_{\xi \beta \kappa}{h_{\tau \epsilon}} \nn \\ && + (\frac{67}{576} p + \frac{1}{384}) \theta^{\nu \kappa} \theta^{\alpha \beta} \partial^{\mu \tau}{}_{\alpha}{h^{\epsilon \xi}} \partial_{\epsilon \xi \beta}{h_{\tau \kappa}}  - \frac{p}{144} \theta^{\xi \alpha} \theta^{\beta \kappa} \partial^{\mu \nu \gamma}h^{\delta}{}_{\xi} \partial_{\delta \alpha \beta}{h_{\gamma \kappa}}  \nn 
\eea
\bea
&& + (\frac{13}{288} p + \frac{1}{384}) \theta^{\xi \alpha} \theta^{\beta \kappa} \partial^{\mu \nu}{}_{\xi}{h^{\gamma \delta}} \partial_{\gamma \alpha \beta}{h_{\delta \kappa}}  - (\frac{1}{768} p - \frac{1}{768}) \theta^{\xi \alpha} \theta^{\beta \kappa} \partial^{\mu \gamma}{}_{\xi}h^{\delta}{}_{\alpha} \partial^{\nu}{}_{\delta \beta}{h_{\gamma \kappa}}  \nn \\ && - (\frac{1}{16} p + \frac{1}{768}) \theta^{\xi \alpha} \theta^{\beta \kappa} \partial^{\mu \gamma}{}_{\xi}{h^{\nu \delta}} \partial_{\delta \alpha \beta}{h_{\gamma \kappa}}  + (\frac{7}{288} p - \frac{1}{768}) \theta^{\xi \alpha} \theta^{\beta \kappa} \partial^{\nu \gamma}{}_{\xi}{h^{\mu \delta}} \partial_{\delta \alpha \beta}{h_{\gamma \kappa}}  \nn \\ && - ( \frac{13}{288} p + \frac{1}{384}) \theta^{\xi \beta} \theta^{\alpha \kappa} \partial^{\mu \nu}{}_{\xi}{h^{\gamma \delta}} \partial_{\gamma \delta \alpha}{h_{\beta \kappa}}  - \frac{1}{384} \theta^{\xi \beta} \theta^{\alpha \kappa} \partial^{\mu \gamma \delta}{h_{\xi \alpha}} \partial^{\nu}{}_{\beta \kappa}{h_{\gamma \delta}} \nn \\ && - ( \frac{17}{144} p + \frac{1}{384}) \theta^{\xi \beta} \theta^{\alpha \kappa} \partial^{\mu \gamma}{}_{\xi}h^{\delta}{}_{\alpha} \partial_{\delta \beta \kappa}h^{\nu}{}_{\gamma}  - (\frac{23}{576} p - \frac{1}{384}) \theta^{\xi \beta} \theta^{\alpha \kappa} \partial^{\mu \gamma}{}_{\xi}h^{\delta}{}_{\alpha} \partial^{\nu}{}_{\beta \kappa}{h_{\gamma \delta}}  \nn \\ && - ( \frac{23}{288} p - \frac{1}{768})\theta^{\xi \beta} \theta^{\alpha \kappa} \partial^{\mu \gamma}{}_{\xi}h^{\delta}{}_{\alpha} \partial^{\nu}{}_{\delta \beta}{h_{\gamma \kappa}}  - \frac{19p}{576} \theta^{\xi \beta} \theta^{\alpha \kappa} \partial^{\mu}{}_{\xi \alpha}{h^{\gamma \delta}} \partial_{\gamma \beta \kappa}h^{\nu}{}_{\delta} \nn \\ && + \frac{19p}{576} \theta^{\xi \beta} \theta^{\alpha \kappa} \partial^{\mu}{}_{\xi \alpha}{h^{\gamma \delta}} \partial_{\gamma \delta \beta}h^{\nu}{}_{\kappa}  + \frac{23p}{576} \theta^{\xi \beta} \theta^{\alpha \kappa} \partial^{\mu}{}_{\xi \alpha}{h^{\gamma \delta}} \partial^{\nu}{}_{\beta \kappa}{h_{\gamma \delta}} \nn \\ && - \frac{3p}{64} \theta^{\xi \beta} \theta^{\alpha \kappa} \partial^{\mu}{}_{\xi \alpha}{h^{\gamma \delta}} \partial^{\nu}{}_{\gamma \beta}{h_{\delta \kappa}} + \frac{p}{144} \theta^{\xi \beta} \theta^{\alpha \kappa} \partial^{\mu}{}_{\xi \alpha}{h^{\gamma \delta}} \partial^{\nu}{}_{\gamma \delta}{h_{\beta \kappa}} \nn \\ && - \frac{p}{9} \theta^{\xi \beta} \theta^{\alpha \kappa} \partial^{\nu \gamma}{}_{\xi}h^{\delta}{}_{\alpha} \partial_{\delta \beta \kappa}h^{\mu}{}_{\gamma} - (\frac{23}{576} p + \frac{1}{384}) \theta^{\xi \beta} \theta^{\alpha \kappa} \partial^{\nu}{}_{\xi \alpha}{h^{\gamma \delta}} \partial_{\gamma \beta \kappa}h^{\mu}{}_{\delta}  \nn \\ && + (\frac{1}{384} + \frac{23}{576} p) \theta^{\xi \beta} \theta^{\alpha \kappa} \partial^{\nu}{}_{\xi \alpha}{h^{\gamma \delta}} \partial_{\gamma \delta \beta}h^{\mu}{}_{\kappa}  + (\frac{29}{192} p + \frac{1}{384}) \theta^{\xi \beta} \theta^{\alpha \kappa} \partial^{\gamma}{}_{\xi \alpha}{h^{\mu \delta}} \partial_{\delta \beta \kappa}h^{\nu}{}_{\gamma}  \nn \\ && + (\frac{137}{576} p - \frac{1}{768}) \theta^{\xi \kappa} \theta^{\alpha \beta} \partial^{\mu \gamma}{}_{\xi}h^{\delta}{}_{\alpha} \partial^{\nu}{}_{\delta \beta}{h_{\gamma \kappa}}  + (\mu \leftrightarrow \nu)
\eea
\bea
C_2^{\mu \nu} & = & \frac{17p}{144} \theta^{\mu \beta} \theta^{\nu \kappa} \partial^{\sigma \gamma}{h^{\delta \alpha}} \partial_{\delta \alpha \beta \kappa}{h_{\sigma \gamma}} + (\frac{13}{144} p + \frac{1}{768}) \theta^{\nu \alpha} \theta^{\beta \kappa} \partial^{\mu \tau}{h^{\epsilon \xi}} \partial_{\epsilon \xi \alpha \beta}{h_{\tau \kappa}}  \nn \\ && + (\frac{1}{36} p + \frac{1}{768}) \theta^{\nu \beta} \theta^{\alpha \kappa} \partial^{\tau}{}_{\alpha}{h^{\epsilon \xi}} \partial_{\epsilon \xi \beta \kappa}h^{\mu}{}_{\tau}  - (\frac{1}{24} p + \frac{1}{256}) \theta^{\nu \beta} \theta^{\alpha \kappa} \partial^{\tau}{}_{\alpha}{h^{\epsilon \xi}} \partial^{\mu}{}_{\epsilon \beta \kappa}{h_{\tau \xi}}  \nn \\ && - \frac{p}{9} \theta^{\nu \beta} \theta^{\alpha \kappa} \partial^{\tau}{}_{\alpha}{h^{\epsilon \xi}} \partial^{\mu}{}_{\epsilon \xi \beta}{h_{\tau \kappa}} + (\frac{1}{24} p + \frac{1}{256}) \theta^{\nu \beta} \theta^{\alpha \kappa} \partial^{\tau \epsilon}h^{\xi}{}_{\alpha} \partial^{\mu}{}_{\xi \beta \kappa}{h_{\tau \epsilon}}  \nn \\ && + (\frac{1}{8} p + \frac{1}{384}) \theta^{\nu \kappa} \theta^{\alpha \beta} \partial^{\tau}{}_{\alpha}{h^{\epsilon \xi}} \partial^{\mu}{}_{\epsilon \xi \beta}{h_{\tau \kappa}}  + (\frac{59}{576} p - \frac{1}{768}) \theta^{\xi \alpha} \theta^{\beta \kappa} \partial^{\mu \gamma}h^{\delta}{}_{\xi} \partial^{\nu}{}_{\delta \alpha \beta}{h_{\gamma \kappa}}  \nn \\ && + \frac{3p}{64} \theta^{\xi \alpha} \theta^{\beta \kappa} \partial^{\mu}{}_{\xi}{h^{\gamma \delta}} \partial_{\gamma \delta \alpha \beta}h^{\nu}{}_{\kappa} - (\frac{7}{576} p + \frac{1}{768}) \theta^{\xi \alpha} \theta^{\beta \kappa} \partial^{\mu}{}_{\xi}{h^{\gamma \delta}} \partial^{\nu}{}_{\gamma \alpha \beta}{h_{\delta \kappa}}  \nn \\ && + (\frac{25}{192} p - \frac{1}{768}) \theta^{\xi \alpha} \theta^{\beta \kappa} \partial^{\nu \gamma}h^{\delta}{}_{\xi} \partial^{\mu}{}_{\delta \alpha \beta}{h_{\gamma \kappa}}  + \frac{19p}{576} \theta^{\xi \alpha} \theta^{\beta \kappa} \partial^{\nu}{}_{\xi}{h^{\gamma \delta}} \partial_{\gamma \delta \alpha \beta}h^{\mu}{}_{\kappa} \nn \\ && - ( \frac{13}{144} p - \frac{1}{384}) \theta^{\xi \alpha} \theta^{\beta \kappa} \partial^{\gamma}{}_{\xi}{h^{\mu \delta}} \partial^{\nu}{}_{\delta \alpha \beta}{h_{\gamma \kappa}}  - ( \frac{5}{96} p - \frac{1}{768}) \theta^{\xi \alpha} \theta^{\beta \kappa} \partial^{\gamma}{}_{\xi}{h^{\nu \delta}} \partial^{\mu}{}_{\delta \alpha \beta}{h_{\gamma \kappa}}  \nn \\ && + (\frac{1}{36} p - \frac{1}{768}) \theta^{\xi \beta} \theta^{\alpha \kappa} \partial^{\mu \nu}{h^{\gamma \delta}} \partial_{\gamma \delta \xi \alpha}{h_{\beta \kappa}}  - ( \frac{5}{144} p - \frac{1}{768}) \theta^{\xi \beta} \theta^{\alpha \kappa} \partial^{\mu}{}_{\xi}{h^{\gamma \delta}} \partial^{\nu}{}_{\gamma \delta \alpha}{h_{\beta \kappa}}  \nn \\ && + \frac{13p}{288} \theta^{\xi \beta} \theta^{\alpha \kappa} \partial^{\nu}{}_{\xi}{h^{\gamma \delta}} \partial^{\mu}{}_{\gamma \delta \alpha}{h_{\beta \kappa}} - (\frac{7}{144} p + \frac{1}{768}) \theta^{\xi \beta} \theta^{\alpha \kappa} \partial^{\gamma \delta}{h_{\xi \alpha}} \partial^{\mu \nu}{}_{\beta \kappa}{h_{\gamma \delta}}  \nn \\ && - ( \frac{1}{16} p + \frac{1}{768}) \theta^{\xi \beta} \theta^{\alpha \kappa} \partial^{\gamma}{}_{\xi}h^{\delta}{}_{\alpha} \partial^{\mu}{}_{\delta \beta \kappa}h^{\nu}{}_{\gamma}  + (\frac{7}{72} p + \frac{1}{384}) \theta^{\xi \beta} \theta^{\alpha \kappa} \partial^{\gamma}{}_{\xi}h^{\delta}{}_{\alpha} \partial^{\mu \nu}{}_{\beta \kappa}{h_{\gamma \delta}}  \nn \\ && - ( \frac{3}{32} p + \frac{1}{768}) \theta^{\xi \beta} \theta^{\alpha \kappa} \partial^{\gamma}{}_{\xi}h^{\delta}{}_{\alpha} \partial^{\mu \nu}{}_{\delta \beta}{h_{\gamma \kappa}}  - (\frac{7}{144} p + \frac{1}{768}) \theta^{\xi \beta} \theta^{\alpha \kappa} \partial^{\gamma}_{\xi}h^{\delta}{}_{\alpha} \partial^{\nu}{}_{\delta \beta \kappa}h^{\mu}{}_{\gamma}  \nn \\ && - ( \frac{7}{144} p + \frac{1}{384}) \theta^{\xi \beta} \theta^{\alpha \kappa} \partial_{\xi \alpha}{h^{\gamma \delta}} \partial_{\gamma \delta \beta \kappa}{h^{\mu \nu}}  + (\frac{1}{16} p + \frac{1}{768}) \theta^{\xi \beta} \theta^{\alpha \kappa} \partial_{\xi \alpha}{h^{\gamma \delta}} \partial^{\mu}{}_{\gamma \beta \kappa}h^{\nu}{}_{\delta}  \nn \\ && - ( \frac{1}{72} p - \frac{1}{256}) \theta^{\xi \beta} \theta^{\alpha \kappa} \partial_{\xi \alpha}{h^{\gamma \delta}} \partial^{\mu}{}_{\gamma \delta \beta}h^{\nu}{}_{\kappa}  - ( \frac{7}{144} p + \frac{1}{768}) \theta^{\xi \beta} \theta^{\alpha \kappa} \partial_{\xi \alpha}{h^{\gamma \delta}} \partial^{\mu \nu}{}_{\beta \kappa}{h_{\gamma \delta}}  \nn \\ && - ( \frac{1}{72} p + \frac{1}{384}) \theta^{\xi \beta} \theta^{\alpha \kappa} \partial_{\xi \alpha}{h^{\gamma \delta}} \partial^{\mu \nu}{}_{\gamma \beta}{h_{\delta \kappa}}  + (\frac{1}{72} p - \frac{1}{768}) \theta^{\xi \beta} \theta^{\alpha \kappa} \partial_{\xi \alpha}{h^{\gamma \delta}} \partial^{\mu \nu}{}_{\gamma \delta}{h_{\beta \kappa}}  \nn \\ && + (\frac{7}{144} p + \frac{1}{768}) \theta^{\xi \beta} \theta^{\alpha \kappa} \partial_{\xi \alpha}{h^{\gamma \delta}} \partial^{\nu}{}_{\gamma \beta \kappa}h^{\mu}{}_{\delta}  + \frac{1}{768} \theta^{\xi \beta} \theta^{\alpha \kappa} \partial_{\xi \alpha}{h^{\gamma \delta}} \partial^{\nu}{}_{\gamma \delta \beta}h^{\mu}{}_{\kappa} \nn \\ && + (\frac{31}{288} p + \frac{1}{256}) \theta^{\xi \kappa} \theta^{\alpha \beta} \partial^{\gamma}{}_{\xi}h^{\delta}{}_{\alpha} \partial^{\mu \nu}{}_{\delta \beta}{h_{\gamma \kappa}}  - \frac{5p}{64} \theta^{\xi \alpha} \theta^{\beta \kappa} \partial^{\nu}{}_{\xi}{h^{\gamma \delta}} \partial^{\mu}{}_{\gamma \alpha \beta}{h_{\delta \kappa}} 
+ (\mu \leftrightarrow \nu)
\eea
where we introduced the compact notation $\partial_{\mu \nu}=\partial_{\mu} \partial_{\nu}$. 

\end{widetext}

\end{document}